\documentclass[aps,showpacs,preprint,superscriptedaddress]{revtex4}%
\usepackage{graphicx}
\usepackage{amsmath}
\usepackage{amsfonts}
\usepackage{amssymb}
\usepackage{bm}
\usepackage{color}
\usepackage{indentfirst}%

\begin{document}
\title{Thomson scattering from high-temperature high-density plasmas revisited }
\author{Jian Zheng}
\email{jzheng@ustc.edu.cn}
\author{C. X. Yu}
\affiliation{CAS Key Laboratory of Basic Plasma Physics, University of Science and
Technology of China, Hefei, Anhui 230026, People's Republic of China}
\affiliation{Department of Modern Physics, University of Science and Technology of China,
Hefei, Anhui 230026, People's Republic of China}

\begin{abstract}
The theory of Thomson scattering from high-temperature
high-density plasmas is revisited from the view point of plasma
fluctuation theory. Three subtle effects are addressed with a
unified theory. The first is the correction of the first order of
$v/c$, where $v$ is the particle velocity and $c$ is the light
speed, the second is the plasma dielectric effect, and the third
is the finite scattering volume effect. When the plasma density is
high, the first effect is very significant in inferring plasma
parameters from the scattering spectra off electron plasma waves.
The second is also be notable but less significant. When the size
of the scattering volume is much larger than the probe wavelength,
the third is negligible.
\end{abstract}

 \maketitle

\section{Introduction}

Thomson scattering is widely used as a key diagnostic tool to
characterize laser-produced plasmas in relevance to inertial
confinement fusion in
recent years~\cite{LaFontaine1994,Glenzer1996,Glenzer1997,Glenzer1999,%
Bai2001,Wang2005,Yu2005,Froula2005,Froula2007}. In comparison with
the plasmas in the field of magnetic confinement fusion,
laser-produced plasmas have some unique properties, such much
higher electron density, much greater parameter gradients, more
frequent Coulomb collisions, and super-Gaussian electron velocity
distribution due to strong inverse bremsstrahlung heating, etc.
The theory of Thomson scattering is developed to include various
effects that may be important for laser-produced plasmas, such as
super-Gaussian electron velocity
distribution~\cite{Zheng1997,Liu2002}, Coulomb
collisions~\cite{Myatt1998,Zheng1999}, plasma
inhomogeneity~\cite{Rozmus2000,Wang2005b}. However, there are
still some subtle effects neglected in inferring plasma parameters
from Thomson scattering spectra. For example, the correction in
the first order of $v/c$ to the scattering spectrum (where $v$ and
$c$ are the electron and light speeds,
respectively)~\cite{Pappert}, the finite scattering volume
effect~\cite{Pechacek1967} and the plasma dielectric
effect~\cite{Sitenko1967}. These effects may also be important for
analyzing the data of Thomson scattering from laser-produced
plasmas. In this article, we revisit the theory of Thomson
scattering from high-temperature high-density plasmas, and give a
unified theory to address these effects from the view point of
fluctuation theory. The correction term of the first order of
$v/c$ that we obtain in this article is a little different from
the result given by Sheffield in his well-known
monograph~\cite{Sheffield1975}. We can see that the correction due
to the finite $v/c$ is very significant in analyzing the
scattering spectra from thermal electron plasma waves. The plasma
dielectric effect is also notable in theory. When the size of the
scattering volume is much larger than the probe wavelength, the
finite scattering volume effect is negligible.

The article is organized as follows. In Sect. 2, we derive basic
equations of Thomson scattering. We then discuss the spectral
power of Thomson scattering with the correction of the first order
of $v/c$ in Sect. 3. The plasma dielectric effect is addressed in
Sect. 4, and the finite scattering volume effect is discussed in
Sect. 5. Finally, we make a summary in Sect. 6.

\section{Basic equations of Thosmon scattering}

We start our calculations from the electric and magnetic fields emitted from
an accelerated electrons~\cite{Landau1975}%
\begin{align*}
\mathbf{E}(\mathbf{R},\mathbf{r},t) &  =-\frac{e}{c^{2}|\mathbf{R}%
-\mathbf{r}(\tau)|[1-\mathbf{n}\cdot\mathbf{v}(\tau)/c]^{3}}\mathbf{n}%
\times\{[\mathbf{n}-\mathbf{v}(\tau)/c]\times\mathbf{\dot{v}}(\tau)\}\\
\mathbf{B}(\mathbf{R},\mathbf{r},t) &  =\mathbf{n}\times\mathbf{E},
\end{align*}
where $-e$ is the electron charge, $\mathbf{\dot{v}}(\tau)=d\mathbf{v}%
(\tau)/d\tau$ is the charge acceleration, $\mathbf{r}$ is the charge
coordinate, $\mathbf{R}$ is the observation coordinate, $\mathbf{n}%
=[\mathbf{R}-\mathbf{r}(\tau)]/|\mathbf{R}-\mathbf{r}(\tau)|$ is the unit
vector from the electron to the observer, and $\tau$ is the retarded time
defined as%
\[
t=\tau+\frac{1}{c}\left\vert \mathbf{R}-\mathbf{r}(\tau)\right\vert .
\]
As usual, we choose the origin of our coordinate system in the
interior of the charge system. In the wave zone, i.e., $R\gg r$,
we can make the following approximations,%
\begin{subequations}
\label{approximation 1}%
\begin{align}
\mathbf{n} &  =\frac{\mathbf{R}}{R},\label{scattering direction}\\
t &  =\tau+\frac{R}{c}-\frac{1}{c}\mathbf{n}\cdot\mathbf{r}(\tau
).\label{retarded time}%
\end{align}
\end{subequations}
The emitted magnetic field in the wave zone can now be
approximately written as
\begin{equation}
\mathbf{B}(\mathbf{R},\mathbf{r},t)=-\frac{e}{c^{2}R[1-\mathbf{n}%
\cdot\mathbf{v}(\tau)/c]}\frac{d}{d\tau}\left[  \frac{\mathbf{v}(\tau
)\times\mathbf{n}}{1-\mathbf{n}\cdot\mathbf{v}(\tau)/c}\right]
.\label{magnetic field}%
\end{equation}
The spectral energy emitted into the solid angle $d\Omega$ is then given by%
\begin{equation}
\frac{d^{2}\mathcal{E}_{\mathbf{n},\omega_{s}}}{d\Omega d\omega_{s}}%
=\frac{e^{2}}{8\pi^{2}c^{3}}\left\vert \int_{-\infty}^{\infty}\frac{d}{d\tau
}\left[  \frac{\mathbf{v}(\tau)\times\mathbf{n}}{1-\mathbf{n}\cdot
\mathbf{v}(\tau)/c}\right]  e^{i\omega_{s}\tau-i(\omega_{s}/c)\mathbf{n}%
\cdot\mathbf{r}(\tau)}d\tau\right\vert ^{2}.\label{radiation energy 1}%
\end{equation}
where $\omega_s$ denotes the frequency of the emitted
electromagnetic waves. In the case that the radiation process
involves many electrons, Eq.~(\ref{radiation energy 1}) should be rewritten as%
\begin{equation}
\frac{d^{2}\mathcal{E}}{d\Omega d\omega_{s}}=\frac{e^{2}}{8\pi^{2}c^{3}%
}\left\vert \sum_{i=1}^{N}\int_{-\infty}^{\infty}\frac{d}{d\tau}\left[
\frac{\mathbf{v}_{i}(\tau)\times\mathbf{n}}{1-\mathbf{n}\cdot\mathbf{v}%
_{i}(\tau)/c}\right]  e^{i\omega_{s}\tau-i(\omega_{s}/c)\mathbf{n}%
\cdot\mathbf{r}_{i}(\tau)}d\tau\right\vert ^{2},\label{radiation energy 2}%
\end{equation}
where $N$ is the total electron number in the system of interest.

In the process of Thomson scattering, an electron is accelerated
under the action of an incident electromagnetic wave and then
emits scattering waves. The intensity of the incident wave is
assumed so weak as not to change the velocity of the electron. For
the sake of simplicity, we further assume that the incident wave
is plane, monochromatic and linearly polarized, i.e., the electric
and magnetic fields of the incident wave are described with
\begin{subequations}
\label{incident field}%
\begin{align}
\mathbf{E}_{0} &  =\mathbf{e}_{0}E_{0}\cos\left(  \mathbf{k}_{0}%
\cdot\mathbf{r}-\omega_{0}t\right)  ,\label{incident electric field}\\
\mathbf{B}_{0} &  =(\mathbf{n}_{0}\times\mathbf{e}_{0})E_{0}\cos\left(
\mathbf{k}_{0}\cdot\mathbf{r}-\omega_{0}t\right)
,\label{incident magnetic field}%
\end{align}
where $\omega_{0}$ and $\mathbf{k}_{0}$ are the frequency and wave
vector of the incident wave, $\mathbf{e}_{0}$ is the polarization
vector of the wave, and $\mathbf{n}_{0}$ is the propagation
direction of the incident wave. In this paper, we are only
interested in the Thomson scattering within the correction to the
first order of $v/c$. Therefore, we neglect those terms which are
proportional to the higher order of $v/c$ in the following
calculations. The acceleration of an electron in the wave fields
of Eq.~(\ref{incident field}) is given by%
\end{subequations}
\begin{equation}
\mathbf{\dot{v}}(\tau)=-\frac{eE_{0}}{m_{e}}\left[  \mathbf{e}_{0}+\frac{1}%
{c}\mathbf{v}\times(\mathbf{n}_{0}\times\mathbf{e}_{0})\right]  \cos
[\mathbf{k}_{0}\cdot\mathbf{r}(\tau)-\omega_{0}\tau]+O[(v/c)^{2}%
].\label{acceleration}%
\end{equation}
With aid of Eq. (\ref{acceleration}), we have%
\begin{align}
\frac{d}{d\tau}\left[  \frac{\mathbf{v}(\tau)\times\mathbf{n}}{1-\mathbf{n}%
\cdot\mathbf{v}(\tau)/c}\right]   &  =-\frac{eE_{0}}{m_{e}}\left\{
(\mathbf{e}_{0}\times\mathbf{n)}+\frac{1}{c}(\mathbf{e}_{0}\times
\mathbf{n})[(\mathbf{n}-\mathbf{n}_{0})\cdot\mathbf{v}]+\frac{1}{c}%
(\mathbf{n}_{0}\times\mathbf{n})(\mathbf{e}_{0}\cdot\mathbf{v})\right.
\nonumber\\
&  \left.  +\frac{1}{c}(\mathbf{e}_{0}\cdot\mathbf{n})(\mathbf{v}%
\times\mathbf{n})\right\}  \cos[\mathbf{k}_{0}\cdot\mathbf{r}(\tau)-\omega
_{0}\tau]+O[(v/c)^{2}].\label{approximation}%
\end{align}
Substituting Eq. (\ref{approximation}) into Eq. (\ref{radiation energy 2}), we
obtain the spectral energy of the Thomson scattering,%
\begin{align}
\frac{d^{2}\mathcal{E}_{\mathbf{n},\omega_{s}}}{d\Omega d\omega_{s}} &
=\frac{1}{\pi}I_{0}r_{e}^{2}\sum_{i,j=1}^{N}\int_{-\infty}^{\infty}d\tau
\int_{-\infty}^{\infty}d\tau^{\prime}\left[  (\mathbf{e}_{0}\times
\mathbf{n})^{2}+\frac{1}{c}\mathbf{a}\cdot\lbrack\mathbf{v}_{i}(\tau
)+\mathbf{v}_{j}(\tau^{\prime})]\right]  \nonumber\\
&  \times\cos[\mathbf{k}_{0}\cdot\mathbf{r}_{i}(\tau)-\omega_{0}\tau
]\cos[\mathbf{k}_{0}\cdot\mathbf{r}_{j}(\tau^{\prime})-\omega_{0}\tau^{\prime
}]e^{i\omega_{s}\tau-i(\omega_{s}/c)\mathbf{n}\cdot\mathbf{r}_{i}(\tau
)}e^{-i\omega_{s}\tau^{\prime}+i(\omega_{s}/c)\mathbf{n}\cdot\mathbf{r}%
_{j}(\tau^{\prime})}\label{radiation energy 3}%
\end{align}
where $I_{0}=cE_{0}^{2}/8\pi$ is the intensity of the incident wave,
$r_{e}=e^{2}/m_{e}c^{2}$ is the classical electron radius, $\mathbf{r}%
_{i}(\tau)$ and $\mathbf{v}_{i}(\tau)$ are the coordinate and velocity of the
$i-$electron, respectively, and the vector $\mathbf{a}$ is defined as%
\begin{equation}
\mathbf{a}=(\mathbf{e}_{0}\times\mathbf{n})^{2}\mathbf{(n}-\mathbf{n}%
_{0}\mathbf{)}-(\mathbf{e}_{0}\cdot\mathbf{n})(\mathbf{n\cdot n}%
_{0})\mathbf{e}_{0}-(\mathbf{e}_{0}\cdot\mathbf{n})^{2}\mathbf{n+}%
(\mathbf{e}_{0}\cdot\mathbf{n})\mathbf{e}_{0}.\label{vector}%
\end{equation}
In the case that $\mathbf{e}_{0}$ is perpendicular to\textbf{ }the scattering
direction $\mathbf{n}$, the vector $\mathbf{a}$ is greatly simplified,%
\begin{equation}
\mathbf{a}=(\mathbf{e}_{0}\times\mathbf{n})^{2}(\mathbf{n}-\mathbf{n}%
_{0}).\label{simplified vector}%
\end{equation}

The electrons in the plasmas can be described with the Klimontovich
distribution function \cite{Lifshitz1981},%
\begin{equation}
F_{e}(\mathbf{r},\mathbf{v},\tau)=\sum_{i=1}^{N}\delta\lbrack\mathbf{r}%
-\mathbf{r}_{i}(\tau)]\delta\lbrack\mathbf{v}-\mathbf{v}_{i}(\tau
)].\label{Klimontovich function 1}%
\end{equation}
Here, it is necessary to assume that $N$ is the number of the
electrons locating within the scattering volume. We just make the
assumption that $N$ is the total electron number of the plasma
within a volume of $V$. In the thermodynamic limit,
$N\rightarrow\infty$, $V\rightarrow\infty$, and the electron
number density $N/V=n_{e}$ is finite. In order to take into
account the effect of finite scattering volume, we introduce a
geometric factor $G(\mathbf{r})$ which is defined as
\begin{equation}
G(\mathbf{r})=\left\{
\begin{array}
[c]{c}%
1,\text{ when }\mathbf{r}\text{ is in }V_{s},\\
0,\text{ when }\mathbf{r}\text{ is not in }V_{s},
\end{array}
\right.  \label{geometry factor}%
\end{equation}
where $V_{s}$ is the scattering volume which is a small part of
the plasma volume $V$. With the geometric factor, only those
electrons locating within the scattering volume $V_{s}$ now
contribute to the scattering power that can be observed. With the
Klimontovich distribution function~(\ref{Klimontovich function 1})
and the geometric factor~(\ref{geometry factor}), the spectral
energy~(\ref{radiation energy 3}) can be
rewritten as%
\begin{align*}
\frac{d^{2}\mathcal{E}_{\mathbf{n},\omega_{s}}}{d\Omega d\omega_{s}} &
=\frac{I_{0}r_{e}^{2}}{\pi}\int_{-\infty}^{\infty}d\tau\int_{-\infty}^{\infty
}d\tau^{\prime}\int d^{3}rd^{3}r^{\prime}d^{3}vd^{3}v^{\prime}\left[
(\mathbf{e}_{0}\times\mathbf{n})^{2}+\frac{1}{c}\mathbf{a}\cdot(\mathbf{v}%
+\mathbf{v}^{\prime})\right]  G(\mathbf{r})G(\mathbf{r}^{\prime})\\
&  \times F_{e}(\mathbf{r},\mathbf{v},\tau)F_{e}(\mathbf{r}^{\prime
},\mathbf{v}^{\prime},\tau^{\prime})\cos(\mathbf{k}_{0}\cdot\mathbf{r}%
-\omega_{0}\tau)\cos(\mathbf{k}_{0}\cdot\mathbf{r}^{\prime}-\omega_{0}%
\tau^{\prime})e^{i\omega_{s}(\tau-\tau^{\prime})-i(\omega_{s}/c)\mathbf{n}%
\cdot(\mathbf{r}-\mathbf{r}^{\prime})}.
\end{align*}
Introducing the variables $\Delta\tau=\tau-\tau^{\prime}$ and $\tau^{\prime
}=\tau^{\prime}$, we rewrite the above equation in the following form,%
\begin{align}
\frac{d^{2}\mathcal{E}_{\mathbf{n},\omega_{s}}}{d\Omega d\omega_{s}} &
=\frac{I_{0}r_{e}^{2}}{\pi}\int_{-\infty}^{\infty}d\tau^{\prime}\int_{-\infty
}^{\infty}d(\Delta\tau)\int d^{3}rd^{3}r^{\prime}d^{3}vd^{3}v^{\prime}\left[
(\mathbf{e}_{0}\times\mathbf{n})^{2}+\frac{1}{c}\mathbf{a}\cdot(\mathbf{v}%
+\mathbf{v}^{\prime})\right]  G(\mathbf{r})G(\mathbf{r}^{\prime})\nonumber\\
&  \times F_{e}(\mathbf{r},\mathbf{v},\tau^{\prime}+\Delta\tau)F_{e}%
(\mathbf{r}^{\prime},\mathbf{v}^{\prime},\tau^{\prime})e^{i\omega_{s}%
\Delta\tau-i(\omega_{s}/c)\mathbf{n}\cdot(\mathbf{r}-\mathbf{r}^{\prime}%
)}\nonumber\\
&  \times\cos[\mathbf{k}_{0}\cdot\mathbf{r}-\omega_{0}(\tau^{\prime}%
+\Delta\tau)]\cos(\mathbf{k}_{0}\cdot\mathbf{r}^{\prime}-\omega_{0}%
\tau^{\prime}).\label{radiation energy 4}%
\end{align}
In many experiments, it is power but not energy that is actually measured. At
the observation point, the relation between the energy and power is given by
the following relation,%
\begin{equation}
\frac{d^{2}\mathcal{E}_{\mathbf{n},\omega_{s}}}{d\Omega d\omega_{s}}%
=\int_{-\infty}^{\infty}\frac{d^{2}P_{\mathbf{n},\omega_{s}}}{d\Omega
d\omega_{s}}dt,\label{energy and power}%
\end{equation}
where $dt$ is the time interval between arrival at the observer of
signals emitted by the source at an interval $d\tau^{\prime}$.
Comparing Eq.~(\ref{energy and power}) with Eq.~(\ref{radiation
energy 4}), and making use of
the relation,%
\begin{equation}
dt=(1-\mathbf{n}\cdot\mathbf{v}^{\prime}/c)d\tau^{\prime}%
,\label{time relation}%
\end{equation}
we obtain the observed spectral power,%
\begin{align}
\frac{d^{2}P_{\mathbf{n},\omega_{s}}}{d\Omega d\omega_{s}} &  =\frac
{I_{0}r_{e}^{2}}{\pi}\int_{-\infty}^{\infty}d(\Delta\tau)\int d^{3}%
rd^{3}r^{\prime}d^{3}vd^{3}v^{\prime}\left[  (\mathbf{e}_{0}\times
\mathbf{n})^{2}+\frac{1}{c}\mathbf{a}\cdot(\mathbf{v}+\mathbf{v}^{\prime
})+\frac{1}{c}(\mathbf{e}_{0}\times\mathbf{n})^{2}(\mathbf{n}\cdot
\mathbf{v}^{\prime})\right]  \nonumber\\
&  \times G(\mathbf{r})G(\mathbf{r}^{\prime})F_{e}(\mathbf{r},\mathbf{v}%
,\tau^{\prime}+\Delta\tau)F_{e}(\mathbf{r}^{\prime},\mathbf{v}^{\prime}%
,\tau^{\prime})e^{i\omega_{s}\Delta\tau-i(\omega_{s}/c)\mathbf{n}%
\cdot(\mathbf{r}-\mathbf{r}^{\prime})}\nonumber\\
&  \times\cos[\mathbf{k}_{0}\cdot\mathbf{r}-\omega_{0}(\tau^{\prime}%
+\Delta\tau)]\cos[\mathbf{k}_{0}\cdot\mathbf{r}^{\prime}-\omega_{0}%
\tau^{\prime}].\label{radiation power 0}%
\end{align}

It should be pointed out that the radiation power emitted by a
radiator is not equal to the radiation power observed at a fixed
point. This is because the time interval at the observation point
is not equal to the time interval of the moving radiator due to
the retarded effect of Eq. (\ref{time relation}). The same
situation is also encountered when one calculates the observed
cyclotron radiation power from a moving charge with a parallel
velocity along the external magnetic field~\cite{Landau1975}. In
Thomson scattering, this effect was first explained as finite
transition time effect by Pechacek and
Trivelpiece~\cite{Pechacek1967}, that exists when the scattering
volume is finite. Lately, this explanation was adopted by
Sheffield in his monograph~\cite{Sheffield1975}. We point out here
that this so-called finite transition effect is essentially the
retarded effect. This is the reason why the finite transition
effect does not in fact depend on the size of scattering volume.

The scattering power given by Eq.~(\ref{radiation power 0}) is a
fluctuating quantity. In experiment, the observation time duration
$\Delta T$ is essentially much larger than the time scales of fluctuations
in the plasma. In this sense, what we observe is in fact the time-averaged spectral power,%
\begin{equation}
\frac{d^{2}P_{\mathbf{n},\omega_{s}}}{d\Omega d\omega_{s}}=\frac
{1}{\Delta T}\int_{-\Delta T/2}^{\Delta T/2}\frac{d^{2}P_{\mathbf{n}%
,\omega_{s}}}{d\Omega d\omega_{s}}dt.\label{averaged power}%
\end{equation}
When the plasma is in equilibrium or quasi-equilibrium state, the
time average on the right hand side of Eq.~(\ref{averaged power})
may be replaced with a proper ensemble average~\cite{Evans1969},
i.e.,
\begin{equation}
\frac{d^{2}P_{\mathbf{n},\omega_{s}}}{d\Omega d\omega_{s}} =
\left\langle \frac{d^{2}P_{\mathbf{n},\omega_{s}}}{d\Omega d\omega_{s}%
}\right\rangle ,\label{emsenble average}%
\end{equation}
where $\langle\cdots\rangle$ denotes ensemble average.

We split the Klimontovich function $F_{e}$ into an ensemble-averaged part
$f_{e}$ and a fluctuating part $\delta f_{e}$,%
\[
F_{e}(\mathbf{r},\mathbf{v},\tau)=f_{e}(\mathbf{v,r})+\delta f_{e}%
(\mathbf{r},\mathbf{v},\tau).
\]
where the fluctuation part $\delta f_{e}$ satisfies the condition%
\[
\langle\delta f_{e}\rangle=0.
\]
When the plasma is in equilibrium or quasi-equ1librium state, the
ensemble-averaged part $f_{e}$ is a Maxwellian distribution function,%
\[
f_{e}=\frac{n_{e}}{(2\pi)^{3/2}v_{e}^{3}}\exp\left(  -\frac{v^{2}}{2v_{e}^{2}%
}\right)  ,
\]
where $v_{e}=\sqrt{T_{e}/m_{e}}$ is the thermal velocity of the electrons.
After ensemble averaged, the spectral power is given by%
\begin{align}
\frac{d^{2}P_{\mathbf{n},\omega_{s}}}{d\Omega d\omega_{s}}  &  =\frac
{I_{0}r_{e}^{2}}{\pi}\int_{-\infty}^{\infty}d(\Delta\tau)\int d^{3}%
rd^{3}r^{\prime}d^{3}vd^{3}v^{\prime}\left[  (\mathbf{e}_{0}\times
\mathbf{n})^{2}+\frac{1}{c}\mathbf{a}\cdot(\mathbf{v}+\mathbf{v}^{\prime
})+\frac{1}{c}(\mathbf{e}_{0}\times\mathbf{n})^{2}(\mathbf{n}\cdot
\mathbf{v}^{\prime})\right] \nonumber\\
&  \times G(\mathbf{r})G(\mathbf{r}^{\prime})\langle\delta f_{e}%
(\mathbf{r},\mathbf{v},\tau^{\prime}+\Delta\tau)\delta f_{e}(\mathbf{r}%
^{\prime},\mathbf{v}^{\prime},\tau^{\prime})\rangle e^{i\omega_{s}\Delta
\tau-i(\omega_{s}/c)\mathbf{n}\cdot(\mathbf{r}-\mathbf{r}^{\prime}%
)}\nonumber\\
&  \times\cos[\mathbf{k}_{0}\cdot\mathbf{r}-\omega_{0}(\tau^{\prime}%
+\Delta\tau)]\cos[\mathbf{k}_{0}\cdot\mathbf{r}^{\prime}-\omega_{0}%
\tau^{\prime}]. \label{radiation power 1}%
\end{align}
If the fluctuation properties of the plasma are known, the
spectral power of Thomson scattering is fully determined.

Fluctuations are usually described with the spectral densities of various
correlation functions. With the Fourier component of the fluctuations,%
\begin{equation}
\delta f_{e}(\omega,\mathbf{k},\mathbf{v})=\int e^{i\omega\tau-i\mathbf{k\cdot
r}}\delta f_{e}(\mathbf{r},\mathbf{v},\tau)d\tau d^{3}r,.
\label{Fourier component}%
\end{equation}
the spectral density of the correlation function of a homogeneous
stationary
plasma is defined by~\cite{Lifshitz1981}%
\begin{equation}
\langle\delta f_{e}(\omega,\mathbf{k},\mathbf{v})\delta f_{e}(\omega^{\prime
},\mathbf{k}^{\prime},\mathbf{v}^{\prime})\rangle=(2\pi)^{4}\delta
(\omega+\omega^{\prime})\delta(\mathbf{k}+\mathbf{k}^{\prime})(\delta
f_{e}^{2})_{\omega,\mathbf{k,v},\mathbf{v}^{\prime}}, \label{spectral density}%
\end{equation}
With aid of Eqs.~(\ref{Fourier component}) and (\ref{spectral
density}), the spectral power of Thomson scattering from a
homogeneous stationary plasma is
given by%
\begin{align}
\frac{d^{2}P_{\mathbf{n},\omega_{s}}}{d\omega_{s}d\Omega}  &  =\frac{1}{2\pi
}I_{0}r_{e}^{2}\int\frac{d^{3}q}{(2\pi)^{3}}d^{3}vd^{3}v^{\prime}%
|G(\mathbf{q})|^{2}(\delta f_{e}^{2})_{\mathbf{k}-\mathbf{q},\omega
,\mathbf{v},\mathbf{v}^{\prime}}\nonumber\\
&  \times\left[  (\mathbf{e}_{0}\times\mathbf{n})^{2}\mathbf{+}\frac{1}%
{c}(\mathbf{v}+\mathbf{v}^{\prime})\cdot\mathbf{a}+\frac{1}{c}(\mathbf{e}%
_{0}\times\mathbf{n})^{2}(\mathbf{n}\cdot\mathbf{v}^{\prime}\mathbf{)}\right]
, \label{radiation power 2}%
\end{align}
where $\omega$ and $\mathbf{k}$ are the differential frequency and wave vector
defined as
\begin{subequations}
\label{difference}%
\begin{align}
\omega &  =\omega_{s}-\omega_{0},\label{differential frequency}\\
\mathbf{k}  &  =\frac{\omega_{s}}{c}\mathbf{n}-\mathbf{k}_{0},
\label{differential wave vector}%
\end{align}
$G(\mathbf{q})$ is the spatial Fourier component of the geometric factor
$G(\mathbf{r})$,%
\end{subequations}
\begin{equation}
G(\mathbf{q})=\int G(\mathbf{r})e^{-i\mathbf{q.r}}d^{3}r.
\label{geometry factor 1}%
\end{equation}

The second and third terms in the bracket on the right hand side
of Eq.~(\ref{radiation power 2}) are the corrections of the first
order of $v/c$. And
the convolution of $|G(\mathbf{q})|^{2}$ and $(\delta f_{e}^{2})_{\mathbf{k}%
-\mathbf{q},\omega,\mathbf{v},\mathbf{v}^{\prime}}$ is the
consequence of finite scattering volume.

\section{First order corrections}

We first discuss the corrections to the first order of $v/c.$ For
the sake of simplicity, we assume that the scattering volume is
infinite. With this
assumption, we have%
\begin{align*}
G(\mathbf{q}) &  =(2\pi)^{3}\delta(\mathbf{q}),\\
\left\vert G(\mathbf{q})\right\vert ^{2} &  =\lim_{V_{s}\rightarrow\infty
}(2\pi)^{3}\delta(\mathbf{q})V_{s}.
\end{align*}
Equation (\ref{radiation power 2}) now becomes%
\begin{equation}
\frac{d^{2}P_{\mathbf{n},\omega_{s}}}{d\omega_{s}d\Omega}=\frac{V_{s}}{2\pi
}I_{0}r_{e}^{2}\int d^{3}vd^{3}v^{\prime}(\delta f_{e}^{2})_{\mathbf{k}%
,\omega,\mathbf{v},\mathbf{v}^{\prime}}\left[  (\mathbf{e}_{0}\times
\mathbf{n})^{2}\mathbf{+}\frac{1}{c}(\mathbf{v}+\mathbf{v}^{\prime}%
)\cdot\mathbf{a}+\frac{1}{c}(\mathbf{e}_{0}\times\mathbf{n})^{2}%
(\mathbf{n}\cdot\mathbf{v}^{\prime}\mathbf{)}\right]
.\label{radiation power 3}%
\end{equation}
For a collisionless quasi-equilibrium plasma, the spectral density $(\delta
f_{e}^{2})_{\mathbf{k},\omega,\mathbf{v},\mathbf{v}^{\prime}}$ is already
obtained \cite{Lifshitz1981},%
\begin{align}
(\delta f_{e}^{2})_{\mathbf{k},\omega,\mathbf{v},\mathbf{v}^{\prime}} &
=2\pi\delta(\mathbf{v}-\mathbf{v}^{\prime})f_{e}(\mathbf{v})\delta
(\omega-\mathbf{k}\cdot\mathbf{v})\nonumber\\
&  +\frac{e^{2}(\mathbf{k}\cdot\partial f_{e}/\partial\mathbf{v}%
)(\mathbf{k}\cdot\partial f_{e}/\partial\mathbf{v}^{\prime})}{m_{e}^{2}%
(\omega-\mathbf{k}\cdot\mathbf{v}+i0)(\omega-\mathbf{k}\cdot\mathbf{v}%
^{\prime}-i0)}\frac{32\pi^{3}}{k^{4}\left\vert \epsilon_{l}(\omega
,k)\right\vert ^{2}}\sum_{\alpha=e,i}e_{\alpha}^{2}\int f_{\alpha}%
(\mathbf{v})\delta(\omega-\mathbf{k}\cdot\mathbf{v})d^{3}v\nonumber\\
&  -\frac{8\pi^{2}e^{2}}{k^{2}}\left\{  \frac{[\mathbf{k}\cdot\partial
f_{e}(\mathbf{v})/\partial\mathbf{v}]f_{e}(\mathbf{v}^{\prime})\delta
(\omega-\mathbf{k}\cdot\mathbf{v}^{\prime})}{m_{e}\epsilon_{l}(\omega
,k)(\omega-\mathbf{k}\cdot\mathbf{v}+i0)}+\frac{f_{e}(\mathbf{v}%
)[\mathbf{k}\cdot\partial f_{e}(\mathbf{v}^{\prime})/\partial\mathbf{v}%
^{\prime}]\delta(\omega-\mathbf{k}\cdot\mathbf{v})}{m_{e}\epsilon_{l}^{\ast
}(\omega,k)(\omega-\mathbf{k}\cdot\mathbf{v}^{\prime}-i0)}\right\}
,\label{spectral density 1}%
\end{align}
where $\epsilon_{l}(\omega,k)$ is the longitudinal permittivity of the plasma.
After integrating over the velocities $\mathbf{v}$ and $\mathbf{v}^{\prime}$,
we obtain (seeing Appendix)%
\begin{equation}
\frac{d^{2}P_{\mathbf{n}_{s},\omega_{s}}}{d\omega_{s}d\Omega}=r_{e}^{2}%
I_{0}N_{s}S(k,\omega)\left[  (\mathbf{e}_{0}\times\mathbf{n})^{2}%
+\frac{2\omega}{kc}\frac{\mathbf{k}}{k}\cdot\mathbf{a}+\frac{\omega}%
{kc}(\mathbf{e}_{0}\times\mathbf{n})^{2}\frac{\mathbf{k}}{k}\cdot
\mathbf{n}\right]  ,\label{radiation power 4}%
\end{equation}
where $N_{s}=n_{e}V_{s}$ is the average electron number in the scattering
volume $V_{s}$. Here $S(k,\omega)$ is the normal dynamic form factor of a
collisionless plasma~\cite{Evans1969},%
\begin{equation}
S(k,\omega)=\frac{1}{\sqrt{2\pi}kv_{e}}\left\vert 1-\frac{\chi_{e}(\omega
,k)}{\epsilon_{l}(\omega,k)}\right\vert ^{2}\exp\left(  -\frac{\omega^{2}%
}{2k^{2}v_{e}^{2}}\right)  +\frac{Z}{\sqrt{2\pi}kv_{i}}\left\vert \frac
{\chi_{e}(\omega/k)}{\epsilon_{l}(\omega/k)}\right\vert ^{2}\exp\left(
-\frac{\omega^{2}}{2k^{2}v_{i}^{2}}\right)  ,\label{dynamic form factor}%
\end{equation}
where $\chi_{e}$ is the electron susceptibility and $Z$ is the ion
charge
state. Here the wave number $k$ is given by%
\begin{align*}
k &  =\left\vert \frac{\omega_{0}}{c}(\mathbf{n}-\mathbf{n}_{0})+\frac{\omega
}{c}\mathbf{n}\right\vert \\
&  \approx\frac{\omega_{0}}{c}\sqrt{2(1-\mathbf{n}\cdot\mathbf{n}_{0})}\left(
1+\frac{\omega}{2\omega_{0}}\right)
\end{align*}

The second and third terms on the right hand side of
Eq.~(\ref{radiation energy 4}) can be further simplified in the
case of
$\sqrt{2(1-\mathbf{n}\cdot\mathbf{n}_{0})}\gg\omega/\omega_{0}$,%
\begin{align*}
\frac{\mathbf{k}}{k} &  \approx\frac{(\mathbf{n}-\mathbf{n}_{0})}%
{\sqrt{2(1-\mathbf{n}\cdot\mathbf{n}_{0})}},\\
k &  \approx\frac{\omega_{0}}{c}\sqrt{2(1-\mathbf{n}\cdot\mathbf{n}_{0})}.
\end{align*}
Then we have%
\begin{align*}
\frac{2\omega}{kc}\frac{\mathbf{k}}{k}\cdot\mathbf{a} &  \approx
(\mathbf{e}_{0}\times\mathbf{n})^{2}\frac{2\omega}{\omega_{0}},\\
\frac{\omega}{kc}(\mathbf{e}_{0}\times\mathbf{n})^{2}\frac{\mathbf{k}}{k}%
\cdot\mathbf{n} &  \approx(\mathbf{e}_{0}\times\mathbf{n})^{2}\frac{\omega
}{2\omega_{0}}.
\end{align*}
Eq.~(\ref{radiation power 4}) is then simplified as%
\begin{equation}
\frac{d^{2}P_{\mathbf{n}_{s},\omega_{s}}}{d\omega_{s}d\Omega}=r_{e}^{2}%
I_{0}N_{s}(\mathbf{e}_{0}\times\mathbf{n})^{2}S(k,\omega)\left(
1+\frac{5\omega}{2\omega_{0}}\right)  .\label{radiation power 5}%
\end{equation}
This result is a little different from that given by Sheffield~\cite{Sheffield1975},%
\begin{equation}
\frac{d^{2}P_{\mathbf{n}_{s},\omega_{s}}}{d\omega_{s}d\Omega}=r_{e}^{2}%
I_{0}N_{s}(\mathbf{e}_{0}\times\mathbf{n})^{2}S(k,\omega)\left(
1+\frac{2\omega}{\omega_{0}}\right)  .\label{Sheffield}%
\end{equation}
The difference between Eq.~(\ref{radiation power 5}) and
Eq.~(\ref{Sheffield}) is rather small, only about
$\omega/2\omega_{0}$. From the derivation of Eq.~(\ref{radiation
power 5}), we can see that this difference comes from the relation
(\ref{time relation}) when we calculate the scattering power.
Without the correction due to Eq. (\ref{time relation}), we return
to the result given by Sheffield.

\begin{figure}
\begin{center}
\includegraphics[height=2.1119in,width=2.6844in]%
{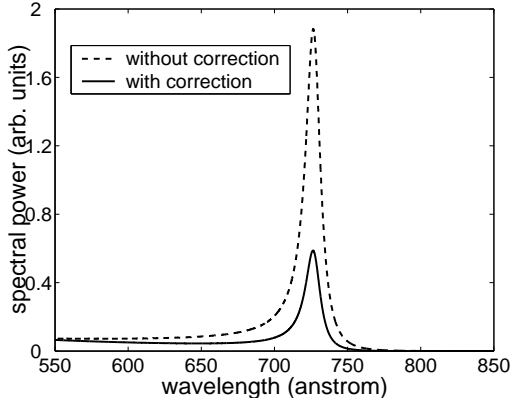}%
\caption{The profile of the spectral powers of Thosmon scattering
with and without the correction, where $T_{e}=2$~keV,
$n_{e}=2.1\times10^{20}$~cm$^{-3}$, the probe wavelength is
$0.5266$~$\mu$m and the scattering angle is
$90^{\circ}$.}%
\label{figure1}%
\end{center}
\end{figure}

Thomson scattering is usually operated in the collective regime in
the measurement of laser-produced plasmas. In most of such
experiments, scattering spectra off thermal ion-acoustic waves are
detected~\cite{LaFontaine1994,Glenzer1996,Glenzer1997,Bai2001,Wang2005,Yu2005,Froula2005}%
. The typical frequency shift of the ion-acoustic features of Thomson
scattering spectra is the order of
\[
\frac{\omega}{\omega_{0}}\sim\sqrt{\frac{ZT_{e}}{Am_{p}}},
\]
where $Z$ is the ion charge state, $A$ is the ion mass number and $m_{p}$ is
the proton mass. For a laser-produced plasma, the electron temperature is
usually lower than $5$ keV. We can see that the correction in the first order
of $v/c$ is negligible. However, scattering spectra from thermal electron
plasma waves are also successfully measured in laser-produced plasmas
\cite{Glenzer1999}. In the experiment performed by Glenzer \textit{et al}.,
the frequency shift $\omega$ of the electron plasma wave feature is very
significant. With a $0.5266$ $\mu$m probe, the spectral maximum locates at
$0.735$ $\mu$m. The correction is then about%
\[
\frac{5}{2}\frac{\omega}{\omega_{0}}=-0.71.
\]
This correction is very large and cannot be neglected. However,
the authors did not include the correction term when fitting the
experimental spectra although many other effects were taken into
account~\cite{Rozmus2000}. Hence, significant errors may be
induced in the process of inferring plasma parameters from the
experimental data. We plot in Fig.~\ref{figure1} the profile of
the scattering power spectrum with the correction (\ref{radiation
power 5}), and compared with that without correction. As seen in
the Fig.~\ref{figure1}, the curve without the correction
overestimated the intensity of the peak corresponding to thermal
electron plasma waves in the plasma. The parameters that we take
in the calculations are the same with those used by Glenzer
\textit{et al}. : $T_{e}=2$ keV, $n_{e}=2.1\times 10^{20}$
cm$^{-3}$, the probe wavelength is $0.5266$ $\mu$m, and the
scattering angle is $90^{\circ}$. Since only the profiles of the
scattering spectra are theoretically fitted, the damping rate of
the electron plasma waves may be overestimated with the
uncorrected scattering theory, leading to
a higher inferred electron temperature.%

\section{Plasma Dielectric Effect}

When plasma density is high, the plasma dielectric effect may also
be significant. In the case that the plasma dielectric effect
cannot be neglected, the intensity of an electromagnetic wave in the plasma is given by%
\[
I=\frac{c}{8\pi}\sqrt{\epsilon(\omega_{0})}E^{2},
\]
where $\epsilon(\omega)=1-\omega^{2}/\omega_{pe}^{2}$ is the
plasma dielectric
constant. And the differential wave vector also changes a little,%
\begin{equation}
\mathbf{k}=\frac{\omega_{s}}{c}\sqrt{\epsilon(\omega_{s})}\mathbf{n}%
-\frac{\omega_{0}}{c}\sqrt{\epsilon(\omega_{0})}\mathbf{n}_{0}%
.\label{differential wave vector 2}%
\end{equation}
The scattering power is now given by%
\begin{equation}
\frac{d^{2}P_{\mathbf{n}_{s},\omega_{s}}}{d\omega_{s}d\Omega}=r_{e}^{2}%
I_{0}N_{s}(\mathbf{e}_{0}\times\mathbf{n})^{2}\sqrt{\frac{\epsilon(\omega
_{s})}{\epsilon(\omega_{0})}}S(k,\omega)\left(  1+\frac{5\omega}{2\omega_{0}%
}\right)  .\label{radiation power 6}%
\end{equation}
Here, the plasma dielectric effect on the correction term is
neglected.

\begin{figure}
[ptb]
\begin{center}
\includegraphics[height=1.868in,width=2.3973in]{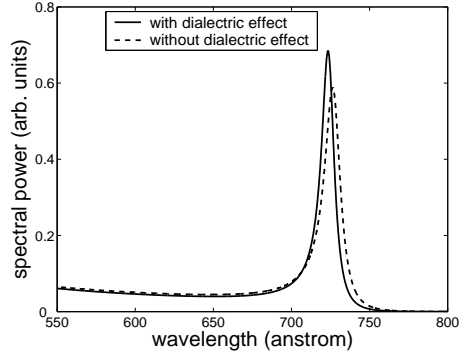}%
\caption{The profile of the spectral power of Thomson scattering
with and without the plasma dielectric effect, where the
parameters is the same
with those used in Fig.~\ref{figure1}}%
\label{figure2}%
\end{center}
\end{figure}

We plot in Fig.~\ref{figure2} the profiles of the features of the
scattering spectrum off an thermal electron plasma wave with and
without the dielectric effect. As seen in Fig. \ref{figure2},
there is some differences when the dielectric effect is taken into
account: the wavelength shift becomes smaller and the intensity
becomes a little higher. The reason is quite understandable. When
the dielectric effect is included, the differential wave vector
becomes a little smaller, leading to a lighter damping and a lower
frequency of the waves. That the factor
$\sqrt{\epsilon(\omega_{s})/\epsilon(\omega_{0})}$ is larger than
$1$ when $\omega<0$ also leads to a little higher intensity. The
plasma dielectric effect becomes more notable when the plasma
density is higher. In the case of $n_{e}/n_{c}$ is about several
percent, the dielectric effect should be included in inferring the
plasma parameters from the features of electron plasma waves.

The dielectric effect on the ion-acoustic features of Thomson
scattering is much less significant. Since $\omega_{s}$ is very
close to the probe frequency $\omega_{0}$ in this case, the
dielectric effect just leads to a minor frequency shift
$\delta\omega/\omega_{ia}\sim0.5(n_{e}/n_{c})$, which is usually
$\lesssim10^{-2}$ in experiment. Therefore, the dielectric effect
can
be neglected in fitting the ion-acoustic feature of a Thomson spectrum.%

\section{Effect of finite scattering volume}

When the scattering volume is small, the geometry factor may become important.
When discussing the finite scattering volume effect, we neglect the correction
terms in the first order of $v/c$. We then have%
\begin{equation}
\frac{d^{2}P_{\mathbf{n}_{s},\omega_{s}}}{d\omega_{s}d\Omega}=n_{e}r_{e}%
^{2}I_{0}(\mathbf{e}_{0}\times\mathbf{n})^{2}\int\frac{d^{3}q}{(2\pi)^{3}%
}|G(\mathbf{q})|^{2}S(|\mathbf{k}-\mathbf{q}|,\omega)
\label{radiation power 7}%
\end{equation}
In the case of $k\gg q$, we can approximately have%
\begin{align*}
\frac{d^{2}P_{\mathbf{n}_{s},\omega_{s}}}{d\omega_{s}d\Omega}  &  \approx
n_{e}r_{e}^{2}I_{0}(\mathbf{e}_{0}\times\mathbf{n})^{2}\int\frac{d^{3}q}%
{(2\pi)^{3}}|G(\mathbf{q})|^{2}\left[  S(k,\omega)-\mathbf{q}\cdot
\frac{\partial}{\partial\mathbf{k}}S(k,\omega)\right] \\
&  \sim N_{s}r_{e}^{2}I_{0}(\mathbf{e}_{0}\times\mathbf{n})^{2}S(k,\omega
)\left[  1+\frac{\Delta q}{k}\right]  .
\end{align*}
Hence $\Delta q$ is the range that $G(\mathbf{q})$ is notably different from
zero. With the assumption that the scattering volume is a cubic with a length
of $L$, $\Delta q\sim2\pi/L$. Then the correction due to the finite scattering
volume effect is about%
\[
\frac{\Delta q}{k}\sim\frac{2\pi}{kL}\sim\frac{\lambda_{0}}{2L\sin(\theta
_{s}/2)},
\]
where $\theta_{s}$ is the scattering angle and $\lambda_{0}$ is the probe
wavelength. For a scattering experiment with a laser wavelength of $0.53$
$\mu$m and the scattering volume size of $50\times50\times50$ $\mu$m$^{3}$ and
scattering angle of $90^{\circ}$, this correction is about $10^{-2}$ and can
be thereby neglected.

\section{Summary}

We revisit the theory of Thomson scattering from high-temperature
high-density plasmas. Three effects are discussed: the correction
due to finite $v/c$, the plasma dielectric effect, and the finite
scattering volume effect. Among these three effects, the
correction due to finite $v/c$ is the most important for analyzing
the scattering spectrum off electron plasma waves. The plasma
dielectric effect is less important but still notable. The finite
scattering volume effect can be neglected if the size of the
scattering volume is about $10^{2}$ of the probe wavelength and
the scattering angle is not very small.

\appendix*

\section{}

We first demonstrate that the normal dynamic form factor of a
collisionless plasma can be obtained from Eq.~(\ref{dynamic form
factor}) with the spectral density (\ref{spectral density 1}).

The electron susceptibility of a collisionless plasma is given by%
\begin{equation}
\chi_{e}(k,\omega)=\frac{4\pi e^{2}}{m_{e}k^{2}}\int\frac{\mathbf{k}%
\cdot\partial f_{e}(\mathbf{v})/\partial\mathbf{v}}{\omega-\mathbf{k}%
\cdot\mathbf{v}+i0}d^{3}v. \label{electron susceptibility}%
\end{equation}
When the distribution function of the electrons is Maxwellian,%
\[
f_{e}=\frac{n_{e}}{(2\pi)^{3/2}v_{e}^{3}}\exp\left(  -\frac{v^{2}}{2v_{e}^{2}%
}\right)  ,
\]
the electron susceptibility can be written as%
\[
\chi_{e}=\frac{1}{k^{2}\lambda_{D}^{2}}\left[  1+\frac{\omega}{\sqrt{2}kv_{e}%
}W\left(  \frac{\omega}{\sqrt{2}kv_{e}}\right)  \right]  ,
\]
where $W(x)$ is the plasma dispersion function. We further introduce the
functions $F_{\alpha}$,%
\[
F_{\alpha}(\omega/k)=\int f_{\alpha}(\mathbf{v})\delta(\omega-\mathbf{k}%
\cdot\mathbf{v})d^{3}v=\frac{n_{\alpha}}{\sqrt{2\pi}kv_{\alpha}}\exp\left(
-\frac{\omega^{2}}{2k^{2}v_{\alpha}^{2}}\right)  .
\]
Then we have%
\[
\int d^{3}vd^{3}v^{\prime}2\pi\delta(\mathbf{v}-\mathbf{v}^{\prime}%
)f_{e}(\mathbf{v})\delta(\omega-\mathbf{k}\cdot\mathbf{v})=2\pi F_{e}%
(\omega/k),
\]%
\begin{align*}
&  \int d^{3}vd^{3}v^{\prime}\frac{e^{2}(\mathbf{k}\cdot\partial
f_{e}/\partial\mathbf{v})(\mathbf{k}\cdot\partial f_{e}/\partial
\mathbf{v}^{\prime})}{m_{e}^{2}(\omega-\mathbf{k}\cdot\mathbf{v}%
+i0)(\omega-\mathbf{k}\cdot\mathbf{v}^{\prime}-i0)}\frac{32\pi^{3}}%
{k^{4}\left\vert \epsilon_{l}(\omega,k)\right\vert ^{2}}\sum_{\alpha
=e,i}e_{\alpha}^{2}\int f_{\alpha}(\mathbf{v})\delta(\omega-\mathbf{k}%
\cdot\mathbf{v})d^{3}v\\
&  =2\pi\left\vert \frac{\chi_{e}(\omega/k)}{\epsilon_{l}(\omega
/k)}\right\vert ^{2}\left[  F_{e}(\omega/k)+Z^{2}F_{i}(\omega/k)\right]  ,
\end{align*}%
\begin{align*}
&  -\frac{8\pi^{2}e^{2}}{k^{2}}\int d^{3}vd^{3}v^{\prime}\left\{
\frac{[\mathbf{k}\cdot\partial f_{e}(\mathbf{v})/\partial\mathbf{v}%
]f_{e}(\mathbf{v}^{\prime})\delta(\omega-\mathbf{k}\cdot\mathbf{v}^{\prime}%
)}{m_{e}\epsilon_{l}(\omega,k)(\omega-\mathbf{k}\cdot\mathbf{v}+i0)}%
+\frac{f_{e}(\mathbf{v})[\mathbf{k}\cdot\partial f_{e}(\mathbf{v}^{\prime
})/\partial\mathbf{v}^{\prime}]\delta(\omega-\mathbf{k}\cdot\mathbf{v})}%
{m_{e}\epsilon_{l}^{\ast}(\omega,k)(\omega-\mathbf{k}\cdot\mathbf{v}^{\prime
}-i0)}\right\} \\
&  =-2\pi\left[  \frac{\chi_{e}}{\epsilon_{l}}+\frac{\chi_{e}^{\ast}}%
{\epsilon_{l}^{\ast}}\right]  F_{e}(\omega/k),
\end{align*}
Combining the above results, we have%
\[
\int d^{3}vd^{3}v^{\prime}(\delta f_{e}^{2})_{\omega,\mathbf{k},\mathbf{v}%
,\mathbf{v}^{\prime}}=2\pi\left\{  \left\vert \frac{1+\chi_{i}(\omega
,k)}{\epsilon_{l}(\omega,k)}\right\vert ^{2}F_{e}(\omega/k)+Z^{2}\left\vert
\frac{\chi_{e}(\omega/k)}{\epsilon_{l}(\omega/k)}\right\vert ^{2}F_{i}%
(\omega/k)\right\}  .
\]
Finally, we obtain the normal dynamic form factor,%
\[
S(k,\omega)=\frac{1}{\sqrt{2\pi}kv_{e}}\left\vert \frac{1+\chi_{i}(\omega
,k)}{\epsilon_{l}(\omega,k)}\right\vert ^{2}\exp\left(  -\frac{\omega^{2}%
}{2k^{2}v_{e}^{2}}\right)  +\frac{Z}{\sqrt{2\pi}kv_{i}}\left\vert \frac
{\chi_{e}(\omega/k)}{\epsilon_{l}(\omega/k)}\right\vert ^{2}\exp\left(
-\frac{\omega^{2}}{2k^{2}v_{i}^{2}}\right)  .
\]

Equation (\ref{radiation power 4}) is verified as follows. It is easy
to demonstrate the following two equations,%
\[
\int\mathbf{v}f_{e}(\mathbf{v})\delta(\omega-\mathbf{k}\cdot\mathbf{v}%
)d^{3}v=\frac{\mathbf{k}}{k}\frac{\omega}{k}F_{e}(\omega/k),
\]%
\[
\int\frac{\mathbf{v}(\mathbf{k}\cdot\partial f_{e}/\partial\mathbf{v})}%
{\omega-\mathbf{k}\cdot\mathbf{v}+i0}d^{3}v=\frac{\mathbf{k}}{k}\frac{\omega
}{k}\int\frac{\mathbf{v}(\mathbf{k}\cdot\partial f_{e}/\partial\mathbf{v}%
)}{\omega-\mathbf{k}\cdot\mathbf{v}+i0}d^{3}v.
\]
With aid of these two equations, we have%
\[
\frac{1}{2\pi n_{e}}\int\mathbf{v}(\delta f_{e}^{2})_{\omega,\mathbf{k}%
,\mathbf{v},\mathbf{v}^{\prime}}d^{3}vd^{3}v^{\prime}=\frac{\mathbf{k}}%
{k}\frac{\omega}{k}S(k,\omega).
\]
Then we obtain Eq.~(\ref{radiation power 4}).

\begin{acknowledgments}
This work is supported by Natural Science Foundation of China
(Nos. 10625523, 10676033).
\end{acknowledgments}

\end{document}